\documentclass[
 reprint,
 superscriptaddress,
nofootinbib,
 amsmath,amssymb,
 aps,
prl,
]{revtex4-2}

\usepackage{graphicx}
\usepackage{dcolumn}
\usepackage{bm}
\usepackage{siunitx}
\usepackage{xspace}
\usepackage{xcolor}
\DeclareSIUnit{\gauss}{G}

\newcommand{\ket}[1]{\ensuremath{\lvert #1 \rangle}\xspace}%
\newcommand{\bra}[1]{\ensuremath{\langle #1 \rvert}\xspace}%
\usepackage[colorlinks=true, citecolor=blue, urlcolor=blue, linkcolor=blue]{hyperref}
\newcommand{\ketbra}[2]{\ensuremath{\lvert #1 \rangle\mkern-2mu\langle #2 \rvert}\xspace}%

\usepackage{cleveref}
\crefname{equation}{Eq.}{Eqs.}
\crefname{observation}{Obs.}{Obs.}
\crefname{figure}{Fig.}{Figs.}

\crefname{corollary}{Corollary}{Corollaries}
\crefname{lemma}{Lemma}{Lemmata}
\crefname{proof}{Proof}{Proofs}
\creflabelformat{proof}{#2proof#3}
\crefname{remark}{Remark}{Remarks}
\crefname{prop}{Proposition}{Propositions}

\usepackage{tikz}
\usetikzlibrary{quantikz2}

\begin{document}

\title{Qutrit entanglement and joint multi-parameter estimation in an optical clock platform}

\author{Ohad Lib$^{\ast,\dagger}$}
\affiliation{Max-Planck-Institut f\"{u}r Quantenoptik, 85748 Garching, Germany}
\affiliation{Munich Center for Quantum Science and Technology (MCQST), 80799 Munich, Germany}

\author{Shuheng Liu$^{\ast}$}
\affiliation{State Key Laboratory for Mesoscopic Physics, School of Physics, Frontiers Science Center for Nano-optoelectronics, Peking University, Beijing 100871, China}
\affiliation{Vienna Center for Quantum Science and Technology, Atominstitut, TU Wien,  1020 Vienna, Austria}
   
\author{Maximilian Ammenwerth}
  \affiliation{Max-Planck-Institut f\"{u}r Quantenoptik, 85748 Garching, Germany}
   \affiliation{Munich Center for Quantum Science and Technology (MCQST), 80799 Munich, Germany}
      
\author{Hendrik Timme}
   \affiliation{Max-Planck-Institut f\"{u}r Quantenoptik, 85748 Garching, Germany}
   \affiliation{Munich Center for Quantum Science and Technology (MCQST), 80799 Munich, Germany}
   \affiliation{Fakult\"{a}t f\"{u}r Physik, Ludwig-Maximilians-Universit\"{a}t, 80799 Munich, Germany}

\author{Shijia Sun}
   \affiliation{Max-Planck-Institut f\"{u}r Quantenoptik, 85748 Garching, Germany}
   \affiliation{Munich Center for Quantum Science and Technology (MCQST), 80799 Munich, Germany}
   \affiliation{Fakult\"{a}t f\"{u}r Physik, Ludwig-Maximilians-Universit\"{a}t, 80799 Munich, Germany}

\author{Qiongyi He}
\affiliation{State Key Laboratory for Mesoscopic Physics, School of Physics, Frontiers Science Center for Nano-optoelectronics, Peking University, Beijing 100871, China}
\affiliation{Collaborative Innovation Center of Extreme Optics, Shanxi University, Taiyuan, Shanxi 030006, China}
\affiliation{Hefei National Laboratory, Hefei 230088, China}

\author{Marcus Huber}
\affiliation{Vienna Center for Quantum Science and Technology, Atominstitut, TU Wien,  1020 Vienna, Austria}
\affiliation{Institute for Quantum Optics and Quantum Information (IQOQI), Austrian Academy of Sciences, 1090 Vienna, Austria}

\author{Giuseppe Vitagliano$^{\ddagger}$}
\affiliation{Vienna Center for Quantum Science and Technology, Atominstitut, TU Wien,  1020 Vienna, Austria}

\author{Immanuel Bloch}
   \affiliation{Max-Planck-Institut f\"{u}r Quantenoptik, 85748 Garching, Germany}
   \affiliation{Munich Center for Quantum Science and Technology (MCQST), 80799 Munich, Germany}
   \affiliation{Fakult\"{a}t f\"{u}r Physik, Ludwig-Maximilians-Universit\"{a}t, 80799 Munich, Germany}

\author{Johannes Zeiher}
\affiliation{Fakult\"{a}t f\"{u}r Physik, Ludwig-Maximilians-Universit\"{a}t, 80799 Munich, Germany}
\affiliation{Max-Planck-Institut f\"{u}r Quantenoptik, 85748 Garching, Germany}
\affiliation{Munich Center for Quantum Science and Technology (MCQST), 80799 Munich, Germany}

\date{\today}
\begin{abstract}

Quantum metrology harnesses entanglement to improve measurement precision beyond classical limits. While standard protocols rely on two-level qubits to estimate a single parameter, extending them to entangled multi-level qudits enables the optimal simultaneous estimation of multiple parameters within a single probe. However, generating such multi-level entanglement and harnessing it for joint multi-parameter estimation in atomic clocks has remained an outstanding challenge. Here, we experimentally demonstrate genuine qutrit entanglement and joint multi-parameter estimation in an optical clock platform. Leveraging control over the ground state and two fine-structure clock states of $^{88}\text{Sr}$ atoms trapped in triple-magic optical tweezers, we generate a maximally entangled two-qutrit state with a loss-postselected fidelity of $F = 0.85(1)$, certifying genuine multi-level entanglement. Taking advantage of this high-dimensional entanglement, we theoretically construct and experimentally realize an optimal two-qutrit metrological probe state and noise-robust readout circuit to simultaneously estimate injected phases on two optical clock transitions. We observe a joint estimation variance below the ideal individual two-level sensing threshold, and show theoretically that this advantage persists at state-of-the-art atom numbers under circuit-level noise. These results demonstrate the key building blocks towards quantum information science with high-dimensional states encoded in the internal energy levels of neutral atoms.

\end{abstract}
\maketitle
{
\renewcommand{\thefootnote}{\fnsymbol{footnote}}
\footnotetext[1]{These authors contributed equally to this work.}
\footnotetext[2]{Corresponding author: ohad.lib@mpq.mpg.de}
\footnotetext[3]{Corresponding author: giuseppe.vitagliano@tuwien.ac.at}
}


Neutral-atom arrays have rapidly emerged as a leading platform for quantum information science, providing an exceptional combination of scalability~\cite{manetsch2025tweezer,chiu2025continuous,li2025fast,muniz2025repeated} and high-fidelity quantum control~\cite{evered2026high,liu2026high,lin2026,lib2026velocity}. These attributes have positioned atomic systems as prime candidates for quantum simulation~\cite{browaeys2020many}, digital quantum computing~\cite{bluvstein2022quantum,graham2022multi}, and quantum-enhanced metrology~\cite{eckner2023realizing,kaubruegger2025progress}, with recent demonstrations showing below-threshold quantum error correction~\cite{bluvstein2025fault} and atomic clocks operating below the standard quantum limit~\cite{eckner2023realizing,robinson2024direct,cao2024multi,finkelstein2024universal}.

To date, most quantum computing and sensing demonstrations have relied on a qubit encoding in two energy levels. A natural generalization is to encode high-dimensional systems, or qudits, in multiple levels of each atom. Such high-dimensional encodings have been theoretically shown to open new possibilities in quantum simulation~\cite{gorshkov2010two}, quantum error correction~\cite{campbell2012magic,campbell2014enhanced,brock2025quditqec}, and quantum metrology~\cite{Szczykulska2016MultiparameterReview,Liu2020QFIMReview}, with proof-of-concept experiments already performed using photonic~\cite{erhard2020advances} and ionic~\cite{ringbauer2022universal,hrmo2023native} entangled qudits. In the context of atomic clocks, the use of two or more transitions allows for exploring multi-parameter estimation. This provides a promising route for mitigating systematic effects such as blackbody radiation shifts~\cite{yudin2011atomic,safronova2018two} and investigating variations of fundamental physical constants~\cite{godun2014frequency,safronova2018two,safronova2018search}. 

Such perspectives have raised significant interest in exploring multi-level encoding in neutral atom platforms~\cite{burshtein2026robust}, including magic trapping conditions~\cite{yamamoto2016ytterbium}, single qudit control~\cite{omanakuttan2021quantum,lindon2023complete,jia2024architecture,ahmed2025coherent,ammenwerth2025realization}, multi-outcome measurements of all qudit states~\cite{gasferrer2026spin,plassmann2026rapid}, and hyper-entanglement with external degrees of freedom~\cite{shaw2025erasure}. However, despite being crucial for qudit-based quantum information processing and metrology, genuine high-dimensional entanglement between individual neutral atoms, encoded in their internal energy levels, has not been demonstrated. As a result, metrological applications such as joint multi-parameter estimation have not been explored with individually controlled atoms and have so far been limited to atomic ensembles~\cite{kunkel2019simultaneous,ahmed2025coherent,cao2025joint,cooper2025entanglement,Li2026MultiparameterAtomicArray}.
\begin{figure*}[t!]
\centering
\includegraphics[width=1\textwidth]{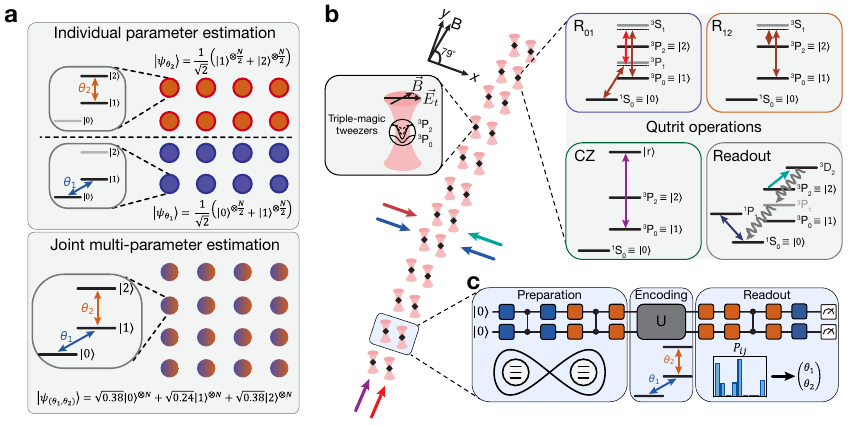}
\caption{\label{fig:1} \textbf{Qutrit entanglement and joint multi-parameter estimation.} \textbf{a} Illustration of individual versus joint estimation of two parameters in an optical clock. In the case of individual estimation, the atoms are split into two groups, each being prepared in a different qubit entangled state to estimate one of the parameters. Alternatively, joint estimation uses qutrit encoding and entanglement across the entire array to estimate both parameters simultaneously. Our experiment is performed for qutrit entangled pairs, for which $N=2$. \textbf{b} Experimental setup consisting of twelve pairs of $^{88}\text{Sr}$ atoms, each encoding a magically trapped qutrit in the ground state and two clock states. Qutrit operations including single-qutrit rotations, entangling gates, and state-resolved readout enable arbitrary global control. Arrows indicate the propagation directions of the control beams with respect to the atomic array, and their corresponding addressed transitions. \textbf{c} Such control allows us to generate and certify qutrit entangled pairs, which are used for joint estimation of two injected phases.}
\end{figure*}

Here, we overcome these limitations and experimentally demonstrate genuine qutrit entanglement and joint two-parameter estimation in an atomic clock platform (Fig.~\ref{fig:1}a,b). We utilize global universal qutrit control over the ground state and two fine-structure (FS) clock states of $^{88}\text{Sr}$ atoms to generate a maximally entangled two-qutrit state with a fidelity of $F=0.85(1)>\frac{2}{3}$, certifying genuine qutrit entanglement~\cite{friis2019entanglement,erhard2020advances}. Using these entangled qutrits, we theoretically and experimentally explore optimal probe states and measurement protocols (Fig.~\ref{fig:1}c) for noise-robust joint estimation of injected phases on both clock transitions. We experimentally estimate both phases jointly with a variance lower than that ideally achievable with perfect individual estimation of each phase~\cite{Humphreys2013QuantumEnhancedMultiplePhase,Chen2019JointRabiFrequencies}. We further show theoretically that the advantage of joint estimation persists at significantly larger atom numbers, enabling future large-scale demonstrations.

Our experiment is based on an all-optical qutrit encoded in the ground state ${^1}\text{S}_0\equiv\ket{0}$ and two FS clock states ${^3}\text{P}_0\equiv\ket{1}$ and ${^3}\text{P}_2, m_J=0\equiv\ket{2}$ of $^{88}\text{Sr}$ atoms~\cite{ammenwerth2025realization}. Twelve pairs of qutrits are trapped in \SI{813}{\nano\meter} optical tweezers, where triple-magic conditions are achieved via angle-tuning of the magnetic field~\cite{ammenwerth2025realization}, ensuring the coherence between the three levels. Arbitrary single-qutrit control is achieved via Raman and three-photon transitions allowing for $R_{12}$ rotations between the FS clock states and $R_{01}$ rotations between the ground and ${^3}\text{P}_0$ clock states~\cite{ammenwerth2025realization,he2025coherent,carman2025collinear,panelli2026microsecond}, respectively, as was demonstrated in our recent work~\cite{ammenwerth2025realization,tao2025universal,lib2026velocity} and is characterized in the Supplementary Information. To entangle pairs of qutrits, we further realize a controlled-Z ($CZ=I-2\ket{11}\bra{11}$) operation via a single-photon transition from the ${^3}\text{P}_0$ state to the $n=47 {}^3\text{S}_1, m_J = -1$ Rydberg state~\cite{tao2025universal,lib2026velocity}. Finally, we realize a loss-detecting measurement protocol which separately detects population in each of the three levels. First, the population in the ground state is destructively measured via fast imaging (\SI{1}{\percent}-level infidelity)~\cite{su2025fast,tao2025universal} on the ${^1}\text{P}_1$ transition. The ${^3}\text{P}_2$ population is then pumped down to the ground state using a selective \SI{496}{\nano\meter} repumper and destructively imaged in the same manner~\cite{tao2025universal}. Finally, the remaining population in ${^3}\text{P}_0$ is pumped to the ground state and measured by a slow lossless image. If no population is detected in any of the three images, the atom is considered lost.

\begin{figure}[t!]
\centering
\includegraphics[width=1\columnwidth]{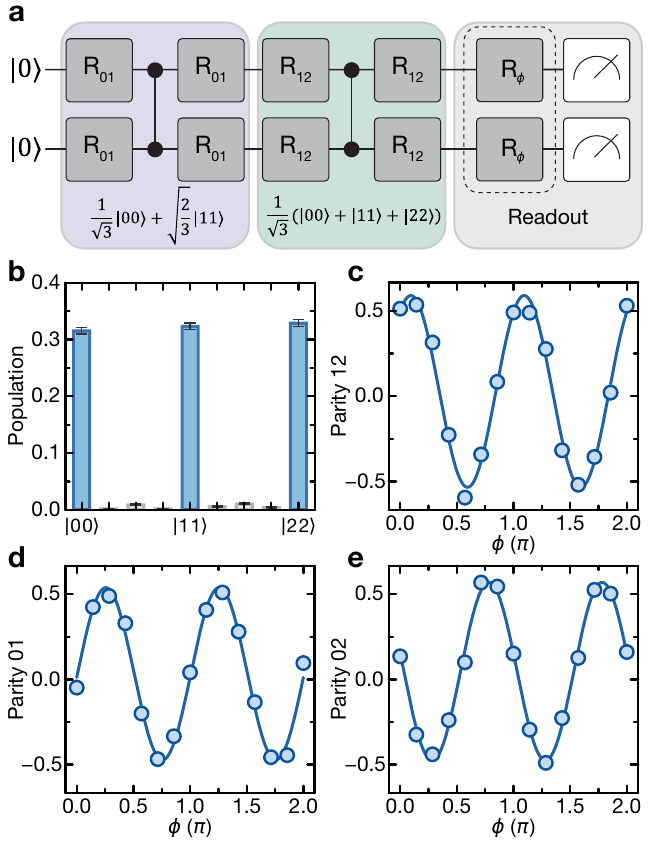}
\caption{\label{fig:2} \textbf{Genuine qutrit entanglement.} \textbf{a} Circuit for creating and benchmarking a qutrit entangled state encoded in the energy levels of two atoms. A phase-scanned $\pi/2$ rotation ($R_{\phi}$) is added when measuring parity oscillations on the different transitions. The exact rotation angles are shown in the Supplementary Information. \textbf{b} State population measurement of the qutrit entangled state. \textbf{c-e} Parity oscillations on the $\ket{1}-\ket{2}$, $\ket{0}-\ket{1}$, $\ket{0}-\ket{2}$ transitions. In \textbf{e}, the $\ket{0}-\ket{2}$ $\pi/2$ pulse is performed using a $\pi$ rotation on the $\ket{1}-\ket{2}$ transition followed by a $\pi/2$ rotation on $\ket{0}-\ket{1}$. The fitted contrasts are $C_{12}=0.56(1)$, $C_{01}=0.51(1)$, and $C_{02}=0.52(2)$ (ideal contrast is $C=\frac{2}{3}$). Combining the measured populations and parity contrasts, we obtain a lower bound for the fidelity of the state $F=\frac{1}{3}(C_{01}+C_{02}+C_{12}+P_{00}+P_{11}+P_{22})=0.85(1)>\frac{2}{3}$, certifying genuine qutrit entanglement.} 
\end{figure}

We combine these magic trapping, control, and measurement techniques to explore qutrit entanglement between pairs of atoms. As a first test-case, we realize the circuit shown in Fig.~\ref{fig:2}a to generate the maximally entangled qutrit state $\ket{\psi}=\frac{1}{\sqrt{3}}(\ket{00}+\ket{11}+\ket{22})$. Using state-resolved detection, we measure the populations of the generated state (Fig.~\ref{fig:2}b) as well as the parity oscillations between all levels (Fig.~\ref{fig:2}c-e). Combining the contrasts of parity oscillations, $C_{ij}$ with the populations in the diagonal states $P_{ii}$, we obtain a fidelity of $F=\frac{1}{3}(C_{01}+C_{02}+C_{12}+P_{00}+P_{11}+P_{22})=0.85(1)>\frac{2}{3}$, clearly certifying genuine qutrit entanglement~\cite{terhal2000schmidt,guhne2009entanglement,friis2019entanglement}. The fidelity is not corrected for any state preparation and measurement errors, aside from discarding cases where either one or both atoms in a pair have been lost (between 15 and \SI{20}{\percent} across the different datasets).

With qutrit encoding and entanglement at hand, we turn to consider multi-parameter estimation across both clock transitions. While single-parameter quantum estimation is well-established~\cite{parisMetrev09,TothApellaniz2014,Demkowicz_Dobrza_ski_2015,DegenReinhardCappellaroRev2017,Pezze2018QuantumMetrologyAtomicEnsembles}, multi-parameter estimation introduces a fundamentally richer and more complex landscape where structural features like commutativity of the parameters~\cite{ragy2016compatibility}, as well as fine-tuning control capabilities, e.g., preparations with high entanglement dimensionality~\cite{Liu2024bounding} and global multi-particle and multi-level measurements, all play a key role for achieving more precise simultaneous estimation of all parameters~\cite{Humphreys2013QuantumEnhancedMultiplePhase,PezzeOptimal2017}. Here, we consider the estimation of two phases $\theta_1$ and $\theta_2$, injected between the $\ket{0}$ and $\ket{1}$ levels and the $\ket{1}$ and $\ket{2}$ levels, respectively. Formally, the corresponding single-qutrit generators are
\begin{equation}
g_1=\frac{1}{\sqrt2}\left(\ketbra{0}{0} - \ketbra{1}{1}\right),
\qquad
g_2=\frac{1}{\sqrt2}\left(\ketbra{1}{1} - \ketbra{2}{2}\right) .
\label{eq:single-particle-generators}
\end{equation}

\begin{figure*}[t!]
\centering
\includegraphics[width=1\textwidth]{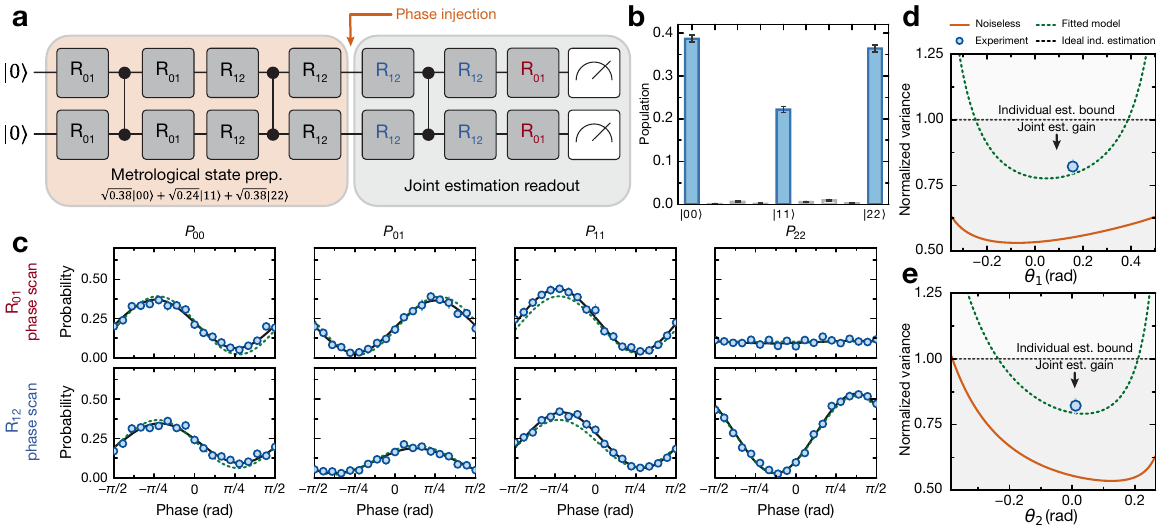}
\caption{\label{fig:3} \textbf{Joint parameter estimation.} \textbf{a} Circuit for joint parameter estimation. The ideal, unbalanced, entangled metrological state is generated, followed by a noise-robust readout circuit to jointly estimate two injected phases (see Supplementary Information for rotation angles). \textbf{b} Population measurement showing the desired imbalance between the qutrit states. \textbf{c} A model allowing for joint phase estimation given probabilities of the nine measurement outcomes is established by scanning the phases of the $R_{01}$ and $R_{12}$ pulses in the readout circuit. The data points are fitted using a cosine-sine fit (black curve). A model taking into account separately measured decoherence in single-qutrit operations shows good agreement with the experimental data (green dashed curve) using only two pulse phase shifts as free parameters for fitting all populations (see Supplementary Information). The probabilities of other outcomes are shown in the Supplementary Information.  \textbf{d,e} For fixed phases of the readout pulses, given the probability model calculated from the data in panel \textbf{c}, we can further obtain the experimental variance in the joint estimation of both phases. This is done through bootstrapping of a loss-postselected dataset at fixed phases (blue data point), showing clear improvement of the achieved variance compared with the ideal individual estimation bound. The resulting variance agrees well with the expected curves calculated with our simplified experimentally calibrated model, plotted as one-dimensional cuts around the measured phases (green dashed curves).}
\end{figure*}

In the standard approach, each parameter can be individually estimated using independent states in a single-parameter estimation protocol. For two-parameter estimation with $N$ atoms per experimental cycle, the parameters can be estimated by entangling half the atoms in the $\ket{\psi_{\theta_1}} = \frac{1}{\sqrt{2}}(\ket{0}^{\otimes \frac{N}{2}}+\ket{1}^{\otimes \frac{N}{2}})$ state, and the other half in the $\ket{\psi_{\theta_2}} = \frac{1}{\sqrt{2}}(\ket{1}^{\otimes \frac{N}{2}}+\ket{2}^{\otimes \frac{N}{2}})$ state~\cite{cao2024multi,finkelstein2024universal}. With optimal measurements, $Y_{\mathrm{ind}}:=(\Delta\theta_1)^2+(\Delta\theta_2)^2\geq \frac{4}{\nu N^2}$ can be achieved for the sum of variances $Y_{\mathrm{ind}}$ using such a split array with qubit entanglement (see Supplementary Information), without changing the effective cycle time of the experiment. Here, $\nu$ denotes the number of repetitions and is hereafter taken to be one unless otherwise specified. Interleaved measurement of each parameter in every second experimental cycle is also possible, yet at the cost of increased Dick-effect noise due to the lower effective measurement rate~\cite{dick1987local}, and is not considered here.

In contrast, using qutrit encoding and entanglement, both parameters can be jointly estimated using all $N$ atoms. As the generators commute, the quantum Cramér-Rao bound can in principle be saturated~\cite{ragy2016compatibility,Humphreys2013QuantumEnhancedMultiplePhase,Szczykulska2016MultiparameterReview,Liu2020QFIMReview}, and the challenge lies in designing and realizing circuits for the optimal state preparation and readout. 

Theoretically, we find that the qutrit-entangled state $\ket{\psi_{(\theta_1,\theta_2)}}=\sqrt{0.38}\ket{0}^{\otimes N}+\sqrt{0.24}\ket{1}^{\otimes N}+\sqrt{0.38}\ket{2}^{\otimes N}$ is optimal among all states and outperforms individual estimation for any number of atoms (see Supplementary Information). The best achievable joint variance is given by:
\begin{equation}
Y_{\mathrm{joint}}=(\Delta\theta_1)^2+(\Delta\theta_2)^2\geq \frac{11+2\sqrt{10}}{9N^2}\approx{0.48}\times Y_{\mathrm{ind}}. 
\label{eq:joint_vs_ind}
\end{equation}

Interestingly, the optimal state is genuinely qutrit-entangled, but has unbalanced populations reflecting the fact that the $\ket{1}$ state is acted on by both generators.

Experimentally, we focus on the two-qutrit case. We find that global operations are sufficient for generating the optimal two-qutrit state $\ket{\psi_{(\theta_1,\theta_2)}}$ and realizing a readout circuit saturating the quantum Cramér-Rao bound of $Y_{\mathrm{joint}}=0.4812$. A viable alternative readout circuit using only single-qutrit gates acting on the entangled probe state would instead achieve $Y_{\mathrm{joint}}=0.5178$, yet with higher sensitivity to gate errors in the state preparation and measurement circuits. Taking into account that $R_{12}$ rotations are better than $R_{01}$ rotations in our system (see Supplementary Information), we optimize a noise-robust readout sequence composed of two $R_{12}$ rotations, a CZ gate, and a single $R_{01}$ gate (Fig.~\ref{fig:3}a). A detailed discussion on noise-robustness and trade-offs in the choice of readout sequence is given in the Supplementary Information.

We experimentally realize the state-preparation circuit, observing the desired unbalanced populations (Fig.~\ref{fig:3}b). Adding the readout part of the circuit, we first separately scan the phases of individual $R_{12}$ and $R_{01}$ rotations in the readout circuit to obtain an experimentally fitted model for the measurement probabilities as a function of $\theta_1$ and $\theta_2$ (Fig.~\ref{fig:3}c). We fit the measurement data of all outcomes using a cosine-sine fit (black curve) and use it to reconstruct a likelihood function for the estimation of both phases (see Supplementary Information). We then extract 
the maximum likelihood estimators (MLEs) for the two parameters, obtained via bootstrapping of a 29,594-shot dataset~\cite{cao2024multi} (see Supplementary Information). From this method, we also calculate the variance in the joint two-parameter estimation, obtaining $\nu[(\Delta\theta_1)^2+(\Delta\theta_2)^2]=0.82(3)<1$, surpassing the ideal individual-estimation bound even though our value includes experimental imperfections while the bound assumes perfect individual estimation.

To understand why we do not saturate the ideal bound of $\approx0.5$, we independently characterize the incoherent damping parameters of the $R_{12}$ and $R_{01}$ gates (see Supplementary Information), and perform a circuit-level simulation of the expected measurement outcomes. Using the independently characterized gates and adding only two phase shifts of the $R_{12}$ and $R_{01}$ readout rotations as free parameters, good agreement is observed with the experimental results (green curve, Fig.~\ref{fig:3}c-e). Further improvements in the fidelity of single-qutrit rotations, for example using feed-forward techniques to mitigate phase noise~\cite{chao2025robust}, should therefore directly benefit the joint parameter estimation. 
\begin{figure}[t]
\centering
\includegraphics[width=1\columnwidth]{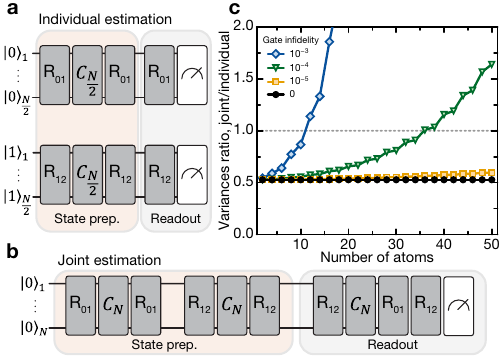}
\caption{\label{fig:4} \textbf{Theoretical noise robustness and scaling.} \textbf{a,b} General circuits for the generation and measurement of metrological states in the cases of individual (\textbf{a}) and joint (\textbf{b}) two-phase estimation using global single-qutrit rotations and entangling operations. The entangling block $C_N$ is defined as $C_N=\prod_{i<j}{\rm CZ}_{ij}$. \textbf{c} For noiseless systems, the joint estimation advantage is kept for an arbitrary number of atoms. Adding circuit-level depolarizing noise for $R_{01}$ and $R_{12}$ rotations and dephasing for individual CZ gates, joint estimation remains favorable for state-of-the-art entangled-state sizes~\cite{cao2024multi,finkelstein2024universal,senoo2026high} with foreseeable gate fidelities.}
\end{figure}

Going beyond qutrit pairs, we theoretically consider the performance of joint estimation for larger qutrit states in the presence of noise. For a large number of atoms, the brute-force circuit optimization used for a pair of qutrits becomes intractable. We instead find a scalable family of circuits for state-preparation and joint two-phase estimation (Fig.~\ref{fig:4}a, b). Interestingly, entangling operations are required to generate a readout that saturates the quantum Cramér-Rao bound in the joint estimation case, but not for individual estimation.

To realistically estimate the advantage of joint estimation, we look at the ratio $Y_{\mathrm{joint}}/Y_{\mathrm{ind}}$ as a function of the number of atoms (Fig.~\ref{fig:4}c). Without noise, the advantage of joint estimation remains constant and persists for an arbitrary number of atoms. When depolarizing and dephasing circuit-level errors are considered for rotations and entangling CZ gates, respectively, the advantage of joint estimation holds up to a finite number of atoms. With high-fidelity operations, joint estimation is expected to outperform individual estimation for state-of-the-art numbers of entangled atoms~\cite{cao2024multi,finkelstein2024universal,senoo2026high}. We expect that future work on constructing lower-depth circuits that require fewer gates, or optimized circuits for a given noise model of the experiment will increase the range of realistic advantage of multi-parameter estimation.
%

%
%

In conclusion, we have experimentally demonstrated joint parameter estimation in an optical clock platform utilizing qutrit entanglement between pairs of neutral atoms. Our first demonstration opens up several directions for future research. Combining large atomic arrays with coherent atom shuttling~\cite{beugnon2007two,lengwenus2010coherent,bluvstein2022quantum} or analog evolution~\cite{senoo2026high} could enable the generation of large multi-partite qutrit entangled states. Besides increasing the precision of multi-parameter estimation, also beyond two phases and including non-commuting generators, such large-scale control over higher-dimensional states could also be useful for realizing novel quantum error correction schemes~\cite{Gottesman1999HigherDimensionalSystems,campbell2012magic,campbell2014enhanced,brock2025quditqec} and the compilation of quantum algorithms~\cite{wang2020qudits}.

In the context of metrology, our proof-of-concept demonstration is currently limited to joint estimation of injected phases, with zero dark time, and without careful characterization of systematic effects. Working at relevant dark times for metrology will require increasing the coherence time in our system, which is limited by polarization gradients across our tweezer array that induce a slight differential light shift between ${^3}\text{P}_2$ and the other states~\cite{ammenwerth2025realization}. This effect becomes especially important at large trap depths, which we use for being in the Lamb-Dicke regime for the $R_{01}$ rotations. Using recoil-free three-photon configurations~\cite{panelli2026microsecond}, improving the polarization homogeneity of the traps, and reducing the trap depth during the dark time should allow future work to explore longer sensing durations.

Finally, we have only considered the estimation of two commuting parameters in qutrits. It will be interesting to explore generalizations to higher-dimensional states with more parameters, especially in the case of non-commuting generators, where the saturation of the quantum Cramér-Rao bound is more challenging~\cite{ragy2016compatibility}. This situation also includes simultaneous estimation of injected phases and noise, realizing a further step toward more noise-resilient quantum circuits.

\begin{acknowledgments}
		We acknowledge funding by the Max Planck Society (MPG), the Deutsche Forschungsgemeinschaft (DFG, German Research Foundation) under Germany's Excellence Strategy--EXC-2111--390814868, and through JST-DFG2024: Japanese-German Joint Call for Proposals on “Quantum Technologies” (Japan-JST-DFG-ASPIRE 2024) under DFG Grant No. 554561799, from the Munich Quantum Valley initiative as part of the High-Tech Agenda Plus of the Bavarian State Government, from the BMFTR through the programs MUNIQC-Atoms and MAQCS, from Quantum Science and Technology-National Science and Technology Major Project (Grants No. 2024ZD0302401 and No. 2021ZD0301500), from National Natural Science Foundation of China (No. 12125402, No. 12534016, and No. 12405005), and from Beijing Natural Science Foundation (Grant No. Z240007).
		This publication has also received funding under Horizon Europe programme HORIZON-CL4-2022-QUANTUM-02-SGA via the project 101113690 (PASQuanS2.1).
		J.Z. acknowledges support from the BMFTR through the program “Quantum technologies---from basic research to market” (SNAQC, Grant No. 13N16265).
        
        O.L. acknowledges support from the Rothschild and CHE Quantum Science and Technology fellowships. S.L. acknowledges the China Postdoctoral Science Foundation (No. 2023M740119). H.T. and M.A. acknowledge funding from the International Max Planck Research School (IMPRS) for Quantum Science and Technology. M.A. acknowledges support through a fellowship from the Hector Fellow Academy.
        This research was funded in whole or in part by the Austrian Science Fund (FWF)  [\href{https://doi.org/10.55776/P35810}{10.55776/P35810}], [\href{https://doi.org/10.55776/P36633}{10.55776/P36633}]. G.V. also acknowledges support from the Grant No. RYC2024-048278-I funded by MCIU/AEI/10.13039/501100011033 and FSE+.
        M.H. acknowledges the Austrian Science Fund (FWF) Grant Nos. 10.55776/COE1 and 10.55776/I6949 and funding from the European Research Council (Consolidator grant ‘Cocoquest’ 101043705).

        Competing interests: J.Z. is a co-founder and shareholder of PlanQC GmbH.

        Data and code availability: The code and data needed to reproduce the main results are publicly available at https://doi.org/10.5281/zenodo.21612351.
	\end{acknowledgments}


\bibliography{bibliography}
\clearpage

\clearpage
\appendix
\section*{Supplementary Information}

\setcounter{figure}{0}
\renewcommand{\thefigure}{S\arabic{figure}}

\renewcommand{\theHfigure}{S\arabic{figure}}

\setcounter{equation}{0}
\renewcommand{\theequation}{S\arabic{equation}}

\renewcommand{\theHequation}{S\arabic{equation}}

\section{Details about the experimental apparatus}

We trap single strontium atoms in optical tweezers at a wavelength of \SI{813}{\nano\meter}. This is a magic wavelength for the ${^1}\text{S}_0$ to ${^3}\text{P}_0$ transition, which can also be made magic with the ${^3}\text{P}_2$ state via angle tuning of the magnetic field with respect to the polarization of the tweezers~\cite{ammenwerth2025realization}. In this experiment, we sort the atoms into twelve pairs ordered in the direction of the UV Rydberg beam, with an intra-pair distance of \SI{2}{\micro\meter} and inter-pair distance of \SI{10}{\micro\meter}. The atoms are cooled near their radial motional ground state ($\bar{n}\approx 0.1$) using sideband cooling on the intercombination line.

All-optical qutrit operations are performed using Raman and three-photon transitions. For $R_{12}$ rotations, we use \SI{11}{\milli\watt} per Raman beam, leading to a Rabi frequency of \SI{280}{\kilo\hertz} with a \SI{12}{\giga\hertz} detuning from ${^3}\text{S}_1$. For $R_{01}$ rotations, we use \SI{2}{\milli\watt} of \SI{688}{\nano\meter} light, \SI{0.5}{\milli\watt} of \SI{689}{\nano\meter} light, and \SI{4}{\milli\watt} of \SI{679}{\nano\meter} light. The detuning from ${^3}\text{P}_1$ is \SI{20}{\mega\hertz}, and the Rabi frequency is \SI{41}{\kilo\hertz}. We choose a trap depth of \SI{500}{\micro\kelvin} which is deep enough for achieving high-fidelity $R_{01}$ rotations while maintaining the coherence of the qutrit for the duration of the circuit.

More details about our experimental apparatus can be found in our recent work which used similar experimental techniques~\cite{ammenwerth2025realization,tao2025universal,lib2026velocity}. 

\subsection{Independent gate characterization}
\label{sec:GateCharacterization}

To model the performance of our joint phase estimation, we characterize the gates comprising the circuit. To keep the model and characterization simple, we make a few simplifying assumptions. First, we take the gates $CZ=I-2\ket{11}\bra{11}$ to be perfect, given their high fidelity in comparison with the single-qutrit operations~\cite{lib2026velocity}. In contrast, we allow for imperfections in the rotation gates $R_{01}$ and $R_{12}$, which are written in matrix form as
\begin{equation}\label{eq:R01MatrixForm}
R_{01}(\vartheta,\phi)=
\begin{pmatrix}
\cos\frac{\vartheta}{2} & -i e^{-i\phi}\sin\frac{\vartheta}{2} & 0\\
-i e^{i\phi}\sin\frac{\vartheta}{2} & \cos\frac{\vartheta}{2} & 0\\
0&0&1
\end{pmatrix},
\end{equation}
\begin{equation}\label{eq:R12MatrixForm}
R_{12}(\vartheta,\phi)=
\begin{pmatrix}
1&0&0\\
0&\cos\frac{\vartheta}{2} & -i e^{-i\phi}\sin\frac{\vartheta}{2}\\
0&-i e^{i\phi}\sin\frac{\vartheta}{2} & \cos\frac{\vartheta}{2}
\end{pmatrix}.
\end{equation}
When the same single-qutrit gate acts simultaneously on both atoms, it is denoted by $R_{ij}^{(2)}=R_{ij}\otimes R_{ij}$.

Instead of full process tomography, we use an effective noise model in which each rotation damps the three Bloch components of its own two-level subspace independently.
For each two-level subspace labeled by $ab\in \{01,12\}$, define
\begin{equation}\label{eq:BlochOperators}
\begin{aligned}
X_{ab} &= \ket{a}\bra{b}+\ket{b}\bra{a},\\
Y_{ab} &= -i\ket{a}\bra{b}+i\ket{b}\bra{a},\\
Z_{ab} &= \ket{a}\bra{a}-\ket{b}\bra{b},
\end{aligned}
\end{equation}
and write the corresponding block of the density matrix as $\rho_{ab}=\frac{1}{2}(I+xX_{ab}+yY_{ab}+zZ_{ab})$ with Bloch vector $\boldsymbol{v}=(1,x,y,z)^{\mathsf T}$. A pulse $R_{ij}(\vartheta,\phi)$ acts on $\boldsymbol{v}$ through the effective transformation
\begin{equation}\label{eq:EffectiveTransferModel}
\begin{aligned}
T(\vartheta,\phi) &= T_{Z}(\phi) T_E T_{R}(\vartheta) T_{Z}(-\phi),\\
T_R(\vartheta) &=
\begin{pmatrix}
1&0&0&0\\
0&1&0&0\\
0&0&\cos\vartheta&-\sin\vartheta\\
0&0&\sin\vartheta&\cos\vartheta
\end{pmatrix},\\
T_Z(\phi) &=
\begin{pmatrix}
1&0&0&0\\
0&\cos\phi&-\sin\phi&0\\
0&\sin\phi&\cos\phi&0\\
0&0&0&1
\end{pmatrix}.
\end{aligned}
\end{equation}
Here $T_E=\operatorname{diag}(1,\eta_x,\eta_y,\eta_z)$. The matrices act from right to left, $T_{R}(\vartheta)$ is the ideal rotation for $\phi=0$, $T_{Z}(\phi)$ rotates the $x$ and $y$ components by $\phi$, and $\eta_x$, $\eta_y$, $\eta_z$ damp the three Bloch components. All pulses of a given type share one parameter set $(\eta_x,\eta_y,\eta_z)$. This describes the measured Bloch-component transformation used for the green comparison curves in \cref{fig:3}.

\begin{figure*}[t]
\centering
\begin{minipage}{0.98\linewidth}
\centering
{\large\bfseries $R_{01}$ repeated-pulse benchmark}\\[6pt]
\begin{quantikz}[
  row sep={0.92cm,between origins},
  column sep=0.48cm
]
\lstick{\scriptsize $Z0:\ \ket{0}$}
& \qw
& \gate[][1.56cm]{\big[R_{01}(\pi/2,0)\big]^n}
& \qw
& \meter{} \rstick[brackets=none]{\scriptsize $P(0)$} \\
\lstick{\scriptsize $X:\ \ket{0}$}
& \gate[][1.32cm]{R_{01}(\pi/2,\pi/2)}
& \gate[][1.56cm]{\big[R_{01}(\pi/2,0)\big]^n}
& \gate[][1.38cm]{R_{01}(\pi/2,3\pi/2)}
& \meter{} \rstick[brackets=none]{\scriptsize $P(0)$} \\
\lstick{\scriptsize $Y:\ \ket{0}$}
& \gate[][1.32cm]{R_{01}(\pi/2,\pi)}
& \gate[][1.56cm]{\big[R_{01}(\pi/2,0)\big]^n}
& \gate[][1.38cm]{R_{01}(\pi/2,0)}
& \meter{} \rstick[brackets=none]{\scriptsize $P(0)$} \\
\lstick{\scriptsize $Z1:\ \ket{1}$}
& \qw
& \gate[][1.56cm]{\big[R_{01}(\pi/2,0)\big]^n}
& \qw
& \meter{} \rstick[brackets=none]{\scriptsize $P(0)$}
\end{quantikz}

\vspace{18pt}

{\large\bfseries $R_{12}$ repeated-pulse benchmark}\\[6pt]
\begin{quantikz}[
  row sep={0.92cm,between origins},
  column sep=0.48cm
]
\lstick{\scriptsize $Z1:\ \ket{1}$}
& \qw
& \gate[][1.56cm]{\big[R_{12}(\pi/2,0)\big]^n}
& \qw
& \meter{} \rstick[brackets=none]{\scriptsize $P(1)$} \\
\lstick{\scriptsize $Z2:\ \ket{2}$}
& \qw
& \gate[][1.56cm]{\big[R_{12}(\pi/2,0)\big]^n}
& \qw
& \meter{} \rstick[brackets=none]{\scriptsize $P(1)$} \\
\lstick{\scriptsize $X:\ \ket{1}$}
& \gate[][1.32cm]{R_{12}(\pi/2,\pi/2)}
& \gate[][1.56cm]{\big[R_{12}(\pi/2,0)\big]^n}
& \gate[][1.38cm]{R_{12}(\pi/2,3\pi/2)}
& \meter{} \rstick[brackets=none]{\scriptsize $P(1)$} \\
\lstick{\scriptsize $Y:\ \ket{1}$}
& \gate[][1.32cm]{R_{12}(\pi/2,\pi)}
& \gate[][1.56cm]{\big[R_{12}(\pi/2,0)\big]^n}
& \gate[][1.38cm]{R_{12}(\pi/2,0)}
& \meter{} \rstick[brackets=none]{\scriptsize $P(1)$}
\end{quantikz}
\end{minipage}
\caption{Single-gate repeated-pulse benchmark sequences. The $Z0/Z1$ and $Z1/Z2$ sequences start in the indicated computational-basis states. In the $X$ and $Y$ sequences, the first pulse prepares the corresponding $X$- or $Y$-basis state and the last pulse maps the same component onto computational-basis readout. The $R_{01}$ curves record $P(0)$, and the $R_{12}$ curves record $P(1)$.}
\label{fig:SingleGateBenchmarkSequences}
\end{figure*}
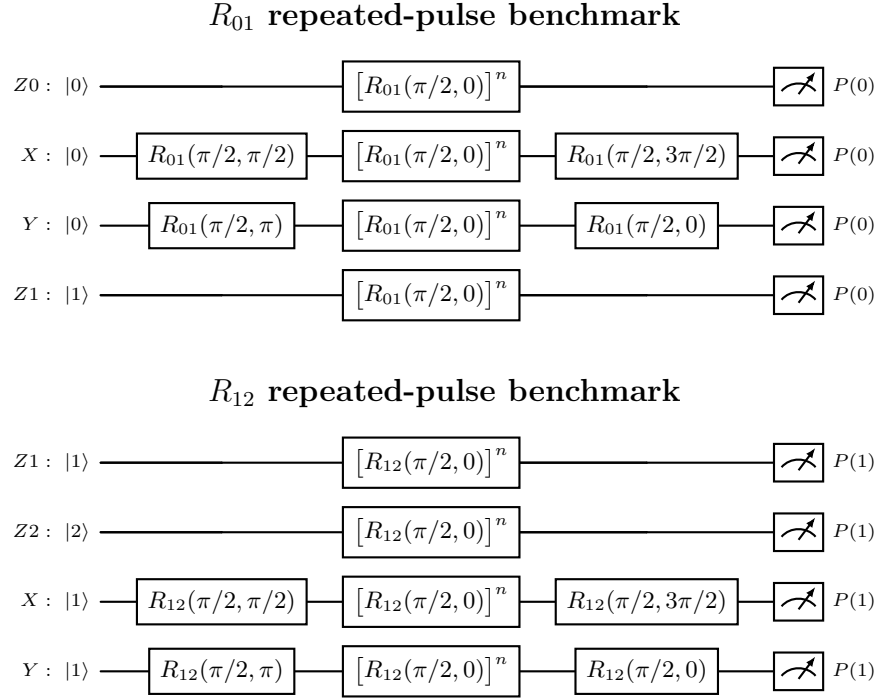

The repeated-pulse sequences used to characterize $R_{01}$ and $R_{12}$ are shown in \cref{fig:SingleGateBenchmarkSequences}. Each sequence contains $n$ repetitions of $R_{ab}(\pi/2,0)$ and ends with a computational-basis population measurement. The $X$ and $Y$ sequences include the preparation and analysis pulses shown in the circuit, which map the corresponding $X$ or $Y$ component onto the measured population. All pulses shown, including the preparation and analysis pulses, enter a joint weighted least-squares fit for each gate, with $0\leq \eta_x,\eta_y,\eta_z\leq 1$.

The fitted parameters are
\begin{equation}\label{eq:FittedDampingParameters}
\begin{aligned}
(\eta_{x,01},\eta_{y,01},\eta_{z,01}) &= (0.9809,0.9341,1.0000),\\
(\eta_{x,12},\eta_{y,12},\eta_{z,12}) &= (0.9868,0.9420,1.0000).
\end{aligned}
\end{equation}
The circuit model used for the green curves uses the six damping parameters together with ideal CZ gates. 

\subsection{Preparation and certification of genuine two-qutrit entanglement}
\label{sec:EntanglementPreparation}

After separately calibrating the gates, we turn to calibrate the full circuits. We start with the preparation of a maximally entangled two-qutrit state
\begin{equation}\label{eq:MaximallyEntangledQutrit}
\ket{\Phi_3}
=\frac{1}{\sqrt{3}}
\left(
\ket{00}+\ket{11}+\ket{22}
\right) ,
\end{equation}
as shown in the circuit in \cref{fig:2}(a). 
All single-qutrit pulses act globally on both atoms. The preparation sequence is
\begin{equation}\label{eq:MESPreparationSequence}
\begin{aligned}
U_{\Phi_3} ={}& R_{12}(\pi/4,0)\cdot CZ\cdot R_{12}(\pi/2,0)\\
&{}\cdot R_{01}(-0.2453\pi,\pi/2)\cdot CZ
\cdot R_{01}(0.5769\pi,\pi/2).
\end{aligned}
\end{equation}

To demonstrate genuine qutrit entanglement, we measure the parity oscillations as in \cref{fig:2}(c)--(e).
These determine $|\rho_{00,11}|$, $|\rho_{11,22}|$, and $|\rho_{00,22}|$, where $\rho_{mn,m'n'}=\bra{mn}\rho\ket{m'n'}$ is a density-matrix element of the prepared state. Appending a global analysis pulse $R_{01}(\pi/2,\phi)$ with scanned phase $\phi$ gives the parity signal
\begin{equation}\label{eq:Parity01}
\Pi_{01}(\phi) = 2\operatorname{Re}(\rho_{01,10})
- 2|\rho_{00,11}|
\cos \left(2\phi+\arg \rho_{00,11}\right).
\end{equation}
The three fitted contrasts therefore satisfy
\begin{equation}\label{eq:ParityContrasts}
\begin{aligned}
C_{01} &= 2|\rho_{00,11}|,\\
C_{12} &= 2|\rho_{11,22}|,\\
C_{02} &= 2|\rho_{00,22}|.
\end{aligned}
\end{equation}
The 1--2 parity is measured in the same way with an $R_{12}(\pi/2,\phi)$ analysis pulse. To measure $|\rho_{00,22}|$, a $\pi$ rotation on the 1--2 transition is applied before the $R_{01}(\pi/2,\phi)$ analysis pulse, as described in \cref{fig:2}(e). For the ideal state $\ket{\Phi_3}$ all three contrasts equal $2/3$.

Combining the fitted contrasts with the measured populations of the three target outcomes gives the fidelity $F=\bra{\Phi_3}\rho\ket{\Phi_3}$,
\begin{equation}\label{eq:Fig2Fidelity}
\begin{aligned}
F &= \frac{1}{3}
\left( P_{00}+P_{11}+P_{22}+C_{01}+C_{02}+C_{12} \right)\\
&= 0.85(1).
\end{aligned}
\end{equation}
The preparation and analysis phases are calibrated so that $\rho_{00,11}$, $\rho_{11,22}$, and $\rho_{00,22}$ have the relative phases expected for the state in \cref{eq:MaximallyEntangledQutrit}. The fidelity is then calculated from their absolute values using \cref{eq:Fig2Fidelity}.

Any two-qutrit state with Schmidt number at most two has an overlap of at most $2/3$ with a maximally entangled two-qutrit state~\cite{terhal2000schmidt}. The measured $F=0.85(1)>2/3$ therefore certifies genuine qutrit entanglement, corresponding to Schmidt number three~\cite{friis2019entanglement,erhard2020advances}. 

\section{Precision bounds and optimal probe}
\label{app:precision_bounds}

Before discussing the phase estimation circuit, we summarize here the theoretical framework used to design and analyze the joint two-parameter estimation protocol. The experiment realizes a qutrit Ramsey interferometer in which the three internal states of each atom are denoted by $\{\ket{0},\ket{1},\ket{2}\}$. The two estimated phases correspond to the relative phases associated with the two adjacent clock transitions. We describe them through the single-qutrit generators
\begin{equation}
\begin{aligned}
g_1 &= \frac{1}{\sqrt{2}}
\left( \ket{0}\bra{0}-\ket{1}\bra{1} \right),\\
g_2 &= \frac{1}{\sqrt{2}}
\left( \ket{1}\bra{1}-\ket{2}\bra{2} \right).
\end{aligned}
\label{eq:SingleQutritGenerators}
\end{equation}
For $N$ atoms, the corresponding collective generators are
\begin{equation}
G_k=\sum_{n=1}^N g_k^{(n)},
\qquad k=1,2,
\label{eq:si-collective-generators}
\end{equation}
where $g_k^{(n)}$ acts on atom $n$. The phase-encoding unitary is
\begin{equation}
U(\theta_1,\theta_2)
=
\exp\left[-i(\theta_1G_1+\theta_2G_2)\right].
\label{eq:CollectiveGeneratorsAndEncoding}
\end{equation}
Since $[G_1,G_2]=0$, the two parameters are compatible at the level of the encoding. Thus, unlike genuinely non-commuting multi-parameter problems, the quantum Cram\'er-Rao bound can be saturated locally by a suitable measurement for pure probe states \cite{holevo2011probabilistic,Szczykulska2016MultiparameterReview,ragy2016compatibility,Liu2020QFIMReview}.

We quantify the performance of the estimation through the cost
\begin{equation}\label{eq:CostDefinition}
Y := (\Delta\theta_1)^2+(\Delta\theta_2)^2,
\end{equation}
which is bounded by the quantum Cram\'er-Rao bound (QCRB) as 
\begin{equation}\label{eq:QuantumCRBCost}
Y \geq \frac{1}{\nu}\operatorname{Tr} ({\cal F}^{-1}),
\end{equation}
where $\nu$ is the number of independent repetitions of the experiment and ${\cal F}$ is the quantum Fisher information matrix. $(\Delta\theta_i)^2$ denotes the
variance constructed from
all $\nu$ repetitions, so that $Y$ scales as $1/\nu$ and the product $\nu Y$ is
the $\nu$-independent figure of merit. All theoretical bounds assume
$\nu = 1$ for simplicity, while the data analysis treats each resolved pair outcome as one
repetition and computes $\nu Y$ with $\nu = N_{\mathrm{shot}}$.

For a pure probe state $\ket{\Psi_0}$ undergoing the unitary encoding in \cref{eq:CollectiveGeneratorsAndEncoding}, one has
\begin{equation}\label{eq:PureStateQFIM}
{\cal F}_{jk}
=4 {\rm Cov}_{\Psi_0}(G_j,G_k),
\end{equation}
where
\begin{equation}
{\rm Cov}_{\Psi_0}(A,B)
=\frac{1}{2}
\bra{\Psi_0}
\{A,B\}
\ket{\Psi_0}
-\bra{\Psi_0}A\ket{\Psi_0}
\bra{\Psi_0}B\ket{\Psi_0} ,
\label{eq:si-generator-covariance}
\end{equation}
and $\{A,B\}=AB+BA$ is the anti-commutator.
Minimizing the QCRB over probe states therefore amounts to optimizing the covariance matrix of the two collective generators.

For the experimentally implemented case $N=2$, this gives
\begin{equation}
Y_{\rm joint}^{\rm opt}
=\frac{11+2\sqrt{10}}{36}
\simeq 0.4812 ,
\label{eq:si-two-qutrit-cost}
\end{equation}
which is the theoretical value used as the ideal joint-estimation benchmark in the main text.

This should be compared with an individual-estimation strategy in which the available atoms are divided into two groups, and each group is used to estimate only one of the two phases with a two-level maximally entangled state. For equal allocation of atoms, the ideal individual-estimation benchmark is
$Y_{\rm ind} \geq 1$. Hence the optimized qutrit protocol improves over the ideal independent two-level strategy exactly by the factor $Y_{\rm joint}^{\rm opt}/Y_{\rm ind} = Y_{\rm joint}^{\rm opt} \simeq 0.4812 $ given in \cref{eq:si-two-qutrit-cost}.
The improvement is closely related to earlier results in multiphase estimation with photonic systems, where estimating several commuting phases jointly can outperform assigning independent resources to each phase \cite{Macchiavello2003OptimalMultiplePhases,Ballester2004MultiplePhases,Humphreys2013QuantumEnhancedMultiplePhase,PezzeOptimal2017,Albarelli2020Perspective}.

In our case, the optimal probe state is of the form
\begin{equation}\label{eq:OptimalProbe}
\ket{\Psi_2^{\rm opt}}
=\lambda_0\ket{0}^{\otimes 2}
+\lambda_1\ket{1}^{\otimes 2}
+\lambda_2\ket{2}^{\otimes 2}.
\end{equation}
where the reduced weight of the $\ket{1}^{\otimes 2}$ component reflects the geometry of the generators: the intermediate state $\ket{1}$ participates in both relative phases. The optimal metrological state has full qutrit support but is deliberately unbalanced and is thus not the maximally entangled qutrit state in \cref{eq:MaximallyEntangledQutrit}.
The latter satisfies instead
\begin{equation}
\operatorname{Tr}\left[{\cal F}^{-1}\right]_{\Phi_3}
=\frac{1}{2},
\end{equation}
i.e., slightly above the optimum.

The state in \cref{eq:OptimalProbe} is already in Schmidt form. Its Schmidt coefficients are
\begin{equation}\label{eq:OptimalSchmidtCoefficients}
\begin{aligned}
\lambda_0=\lambda_2 &= \sqrt{\frac{10-\sqrt{10}}{18}},\\
\lambda_1 &= \sqrt{\frac{\sqrt{10}-1}{9}}.
\end{aligned}
\end{equation}
The state therefore has Schmidt rank three and is genuinely qutrit entangled. This connects the metrological advantage to the use of genuine high-dimensional entanglement, in the same sense that a fidelity larger than $2/3$ with respect to $\ket{\Phi_3}$ certifies Schmidt number three \cite{terhal2000schmidt,guhne2009entanglement,friis2019entanglement,erhard2020advances}.

Numerically constraining the Schmidt rank of the two-qutrit probe gives the following minimal values of $\operatorname{Tr}[{\cal F}^{-1}]$:
\begin{equation}
\begin{split}
r=3: &\qquad 0.4812,\\
r=2: &\qquad 0.6193,\\
r=1: &\qquad 0.9625.
\end{split}
\label{eq:si-schmidt-rank-bounds}
\end{equation}
States with Schmidt rank smaller than three therefore cannot reach the ideal optimum.

\section{Saturation of QCRB with optimal measurements}

We now describe the ideal readout. For a fixed measurement $\{\Pi_x\}$, the outcome probabilities are
\begin{equation}
p(x|\theta_1,\theta_2)
=
\operatorname{Tr}\left[
\Pi_x
U(\theta_1,\theta_2)\rho_{\rm prep}U^\dagger(\theta_1,\theta_2)
\right].
\label{eq:si-probabilities}
\end{equation}
The corresponding classical Fisher information matrix is
\begin{equation}
F_{jk}(\theta_1,\theta_2)
=
\sum_x
\frac{
\partial_{\theta_j}p(x|\theta_1,\theta_2)
\partial_{\theta_k}p(x|\theta_1,\theta_2)
}{
p(x|\theta_1,\theta_2)
}.
\label{eq:si-cfim}
\end{equation}
The classical Cram\'er-Rao bound gives a tighter lower bound on the estimation uncertainty
\begin{equation}
{\rm Cov}(\hat{\theta})
\geq
\frac{1}{\nu}F^{-1}(\theta_1,\theta_2) \geq
\frac{1}{\nu}{\cal F}^{-1} ,
\end{equation}
which, in our case of phases generated by compatible operators, can coincide with the quantum CRB for an optimal measurement. 

A similar statement holds for our cost function
\begin{equation}
Y\geq
\frac{1}{\nu}
\operatorname{Tr}\left[F^{-1}(\theta_1,\theta_2)\right] \geq
\frac{1}{\nu}\operatorname{Tr}\left[{\cal F}^{-1}\right] ,
\label{eq:si-classical-crb}
\end{equation}
with saturation potentially achieved for an optimal measurement.

In particular, in the ideal pure-state model, there exists a locally optimal projective measurement that saturates the quantum Fisher information matrix at the operating point, because the two generators commute and the standard multi-parameter compatibility condition is satisfied \cite{ragy2016compatibility,Szczykulska2016MultiparameterReview,Liu2020QFIMReview}. Experimentally, however, the measurement must be implemented using the available qutrit rotations and Rydberg entangling gates. We therefore optimize the readout directly at the level of the classical Fisher information in \cref{eq:si-cfim}.

For the two-qutrit experiment, the readout consists of a fixed unitary $V_{\rm ro}$ followed by state-resolved measurement in the computational basis. The nine POVM elements are
\begin{equation}
M_{mn}
=
V_{\rm ro}^\dagger \ket{mn}\bra{mn} V_{\rm ro},
\qquad
m,n\in\{0,1,2\}.
\label{eq:si-readout-povm}
\end{equation}
The corresponding probabilities are
\begin{equation}
p_{mn}(\theta_1,\theta_2)
=
\operatorname{Tr}\left[
M_{mn}
U(\theta_1,\theta_2)\rho_{\rm prep}U^\dagger(\theta_1,\theta_2)
\right].
\label{eq:si-readout-probabilities}
\end{equation}
In practice, $V_{\rm ro}$ is chosen from the experimentally available gate set, namely global $R_{01}$ and $R_{12}$ rotations together with controlled-$Z$ gates. 

The readout is optimized for a probe of finite purity, described as the optimal state mixed with global white noise,
\begin{equation}\label{eq:ReadoutDesignWhiteNoiseModel}
\rho_{\eta}^{(2)} = \eta\ket{\Psi_2^{\rm opt}}\bra{\Psi_2^{\rm opt}}
+(1-\eta)\frac{I_9}{9},
\end{equation}
From the measured preparation performance we estimate $\eta = 0.9$, which we use in the optimization below.
The candidate sequences are restricted to the operations available in the experiment. Each contains exactly one global $R_{01}$ rotation, fixed as the final operation, and the preceding operations are drawn from global $R_{12}$ rotations and $CZ$ gates, with adjacent identical operations excluded. 
For every length $L = 1,\ldots,9$ we enumerate the allowed sequences and numerically optimize all rotation angles and phases at $\boldsymbol{\theta} = (0,0)$, taking the cost $\operatorname{Tr}[F^{-1}]$ of \cref{eq:si-classical-crb} as the objective.

\begin{figure}[t]
\centering
\includegraphics[width=\columnwidth]{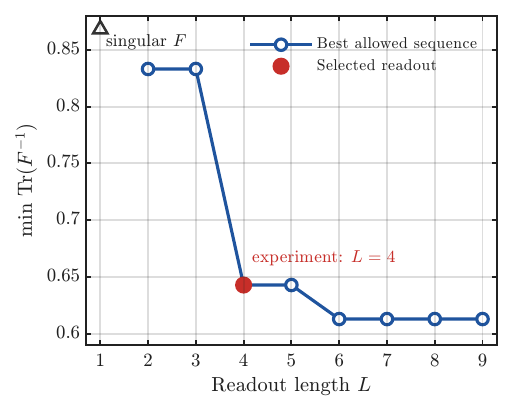}
\caption{Constrained optimization of the two-qutrit readout. Minimum cost $\operatorname{Tr}[F^{-1}]$ at $\boldsymbol{\theta} = 0$ (vertical axis) against the number of operations $L$ in the readout sequence (horizontal axis), evaluated for the probe with $10\%$ white noise in \cref{eq:ReadoutDesignWhiteNoiseModel}. Each candidate sequence ends with a single global $R_{01}$ rotation, preceded by global $R_{12}$ rotations and $CZ$ gates. The angles and phases of every sequence are optimized separately. Lower values correspond to higher precision. The cost drops at $L = 4$, marked by the red point and used in the experiment.}
\label{fig:ReadoutLengthOptimization}
\end{figure}

From the optimized cost in \cref{fig:ReadoutLengthOptimization}, we choose $L = 4$ as a trade-off between the additional errors introduced by more gates and the limited precision gain for $L > 4$.

The selected readout has the structure
\begin{equation}
\begin{aligned}
V_{\mathrm{ro}} ={}& R_{01}(v_3,\phi_3)\cdot R_{12}(v_2,\phi_2)
\cdot CZ\cdot R_{12}(v_1,\phi_1),
\end{aligned}
\label{eq:TwoQutritReadoutStructure}
\end{equation}
where $v_i$ and $\phi_i$ are the angle and phase of the $i$th global single-qutrit rotation, each applied to both atoms, followed by detection in the computational basis. The corresponding parameters are given in \cref{eq:ScannedReadout} at $\delta_{01} = \delta_{12} = 0$.

\begin{figure}[t]
\centering
\includegraphics[width=\columnwidth]{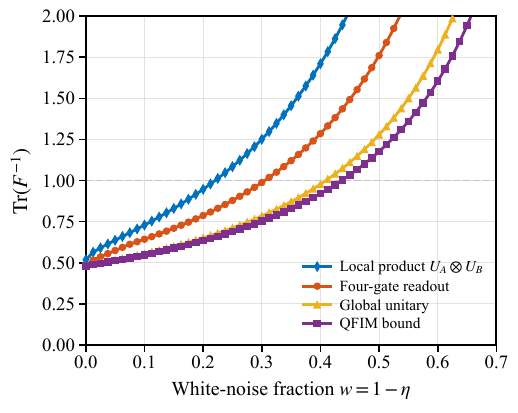}
\caption{$\operatorname{Tr}(F^{-1})$ bounds for the two-qutrit probe under global white noise. At each noise fraction $w=1-\eta$, the blue, orange, and yellow curves give the optimized $\operatorname{Tr}(F^{-1})$ for local-product unitaries $U_A\otimes U_B$, the four-gate readout of \cref{eq:TwoQutritReadoutStructure}, and a global unitary, respectively. The purple curve is the QFIM bound. The horizontal dashed line marks the noiseless individual-estimation bound $Y_{\rm ind}=1$.}
\label{fig:LocalReadoutComparison}
\end{figure}

To compare the selected readout with broader measurement classes, \cref{fig:LocalReadoutComparison} shows their optimized costs $\operatorname{Tr}(F^{-1})$ as a function of the global white-noise fraction $w=1-\eta$.
The three classes compared are product of local unitaries, the four-gate readout of \cref{eq:TwoQutritReadoutStructure}, and a global unitary. In the noiseless limit, the four-gate readout already reaches the quantum bound of $0.4812$, whereas the best local measurement $U_A\otimes U_B$ gives $0.5178$. At $w=0.1$, the four-gate readout gives $0.6430$.

\subsection{Phase-estimation sequence calibration}
\label{sec:SequenceCalibration}

We now describe the experimental calibration of the full metrology circuit, as shown in \cref{fig:3}(a). 
The first set of gates creates the unbalanced entangled state between the $\ket{0}$ and $\ket{1}$ states. We scan the phase of the last $\pi/4$ gate to maximize $P_{00}+P_{11}$, and the duration of the first gate to achieve the correct imbalance between the populations. We then add the second set of gates that extend the entanglement to the $\ket{2}$ state, and repeat the same process, first maximizing $P_{11}+P_{22}$ and then fine-tuning the population imbalance.
Afterwards, we add the rotation gates one by one and correct for relative accumulated phases between them by scanning the phase of the added gate and comparing with the expected oscillations. Such phases arise from single- or two-qutrit gates inducing relative phases between states, which can in principle be calibrated in a sequence-independent way in the future.

The target preparation unitary of the probe state can be written as
\begin{equation}
\begin{aligned}
U_{\mathrm{prep}} ={}& R_{12}(0.2479\pi,0)\cdot CZ
\cdot R_{12}(0.5513\pi,0)\\
&{}\cdot R_{01}(-0.2476\pi,\pi/2)\cdot CZ
\cdot R_{01}(0.5547\pi,\pi/2).
\end{aligned}
\label{eq:ProbePreparation}
\end{equation}
The two phase scans add $\delta_{01}$ to the readout $R_{01}$ phase and $\delta_{12}$ to both readout $R_{12}$ phases, giving the readout unitary
\begin{equation}
\begin{aligned}
V_{\mathrm{ro}}(\delta_{01},\delta_{12}) ={}& R_{01}(1.5708,1.7715+\delta_{01})\\
&{}\cdot R_{12}(1.5708,1.1873+\delta_{12})
\cdot CZ\\
&{}\cdot R_{12}(4.7124,2.7581+\delta_{12}),
\end{aligned}
\label{eq:ScannedReadout}
\end{equation}
followed by detection in the computational basis.

One phase scan varies $\delta_{01}$ at $\delta_{12}=0$, and the other varies $\delta_{12}$ at $\delta_{01}=0$. For one qutrit, the phase encoding is diagonal, and the rotation is given by 
\begin{equation}
\begin{aligned}
R_{01}(\vartheta,\phi)u(\theta_1,\theta_2)
&= u(\theta_1,\theta_2)R_{01}\left(
\vartheta,\phi-\sqrt{2}\theta_1+\frac{\theta_2}{\sqrt{2}}\right),\\
R_{12}(\vartheta,\phi)u(\theta_1,\theta_2)
&= u(\theta_1,\theta_2)R_{12}\left(
\vartheta,\phi+\frac{\theta_1}{\sqrt{2}}-\sqrt{2}\theta_2\right) ,
\end{aligned}
\label{eq:EncodingRotationExchange}
\end{equation}
with $u(\theta_1,\theta_2)
= \operatorname{diag}\left(
e^{-i\theta_1/\sqrt{2}},
e^{i(\theta_1-\theta_2)/\sqrt{2}},
e^{i\theta_2/\sqrt{2}}\right)$.

For two qutrits, the encoding is $u^{\otimes 2}$, where $u=u(\theta_1,\theta_2)$. Since $u^{\otimes 2}$ commutes with $CZ$, applying \cref{eq:EncodingRotationExchange} successively to the readout sequence gives
\begin{equation}
\begin{aligned}
V_{\mathrm{ro}}(0,0)u^{\otimes 2}
&= u^{\otimes 2}V_{\mathrm{ro}}(\delta_{01},\delta_{12}).
\end{aligned}
\label{eq:ReadoutEncodingEquivalence}
\end{equation}
It follows that
\begin{equation}
\begin{pmatrix}
\delta_{01}\\
\delta_{12}
\end{pmatrix} = \begin{pmatrix}
-\sqrt{2} & 1/\sqrt{2}\\
1/\sqrt{2} & -\sqrt{2}
\end{pmatrix}
\begin{pmatrix}
\theta_1\\
\theta_2
\end{pmatrix}.
\label{eq:ScanToParameterMap}
\end{equation}

The probability model is calibrated from the separate $R_{01}$ and $R_{12}$ phase scans described earlier. The calibration retains records in which both atoms have a resolved qutrit outcome. The nine outcome frequencies are averaged with equal weight over the atom pairs.

For $R\in \{01,12\}$, the response of every outcome $m,n\in \{0,1,2\}$ to the scanned phase $\delta$ is fitted with the same cosine-sine form
\begin{equation}
f_{R,mn}(\delta) = a_{R,mn}+b_{R,mn}\cos (2\delta)+c_{R,mn}\sin (2\delta),
\label{eq:ScanResponseModel}
\end{equation}
where the coefficients are obtained by weighted least-squares fits, with the inverse squares of the standard errors of the mean across atom pairs used as weights.
Evaluating the two fitted functions at $\delta=0$ gives $f_{01,mn}(0)$ and $f_{12,mn}(0)$ for outcome $mn$. Their inverse-variance weighted average defines the unnormalized reference value $q_{mn}^{(0)}$, after which
\begin{equation}\label{eq:ScanProbabilityModel}
\begin{aligned}
q_{mn}(\delta_{01},\delta_{12}) &= q_{mn}^{(0)}
+f_{01,mn}(\delta_{01})-f_{01,mn}(0)\\
&\quad +f_{12,mn}(\delta_{12})-f_{12,mn}(0),\\
p_{mn}(\delta_{01},\delta_{12}) &= \frac{q_{mn}(\delta_{01},\delta_{12})}
{\sum_{a,b=0}^{2}q_{ab}(\delta_{01},\delta_{12})}.
\end{aligned}
\end{equation}
The scanned pulse phases are related to the estimated phases by \cref{eq:ScanToParameterMap}. The fitted responses and normalization in \cref{eq:ScanResponseModel,eq:ScanProbabilityModel} give the probabilities $p_{mn}(\theta_1,\theta_2)$ that are then used to extract the likelihood function and hence the Maximum Likelihood Estimator.

The green dashed curves in \cref{fig:3} test whether the independently characterized rotation errors account for the measured outcome probabilities. 
They are obtained by propagating the full pulse sequence with the six damping parameters of \cref{eq:FittedDampingParameters} and ideal $CZ$ gates, all fixed in advance. 
The only free parameters are two small phase offsets that the calibration described above leaves on the readout pulses, a bias $\phi_b^{01}$ on the final readout $R_{01}$ pulse and a bias $\phi_b^{12}$ shared by the two readout $R_{12}$ pulses. 
A least-squares fit to all nine outcome probabilities of both phase scans gives $\phi_b^{01}\approx -0.1077$ and $\phi_b^{12}\approx -0.0057$. 
Both biases are confined to the green curves, while the likelihood in \cref{eq:MLEDefinition} is evaluated with the scan-derived probabilities of \cref{eq:ScanProbabilityModel}.

\begin{figure*}[t]
\centering
\includegraphics[width=\textwidth]{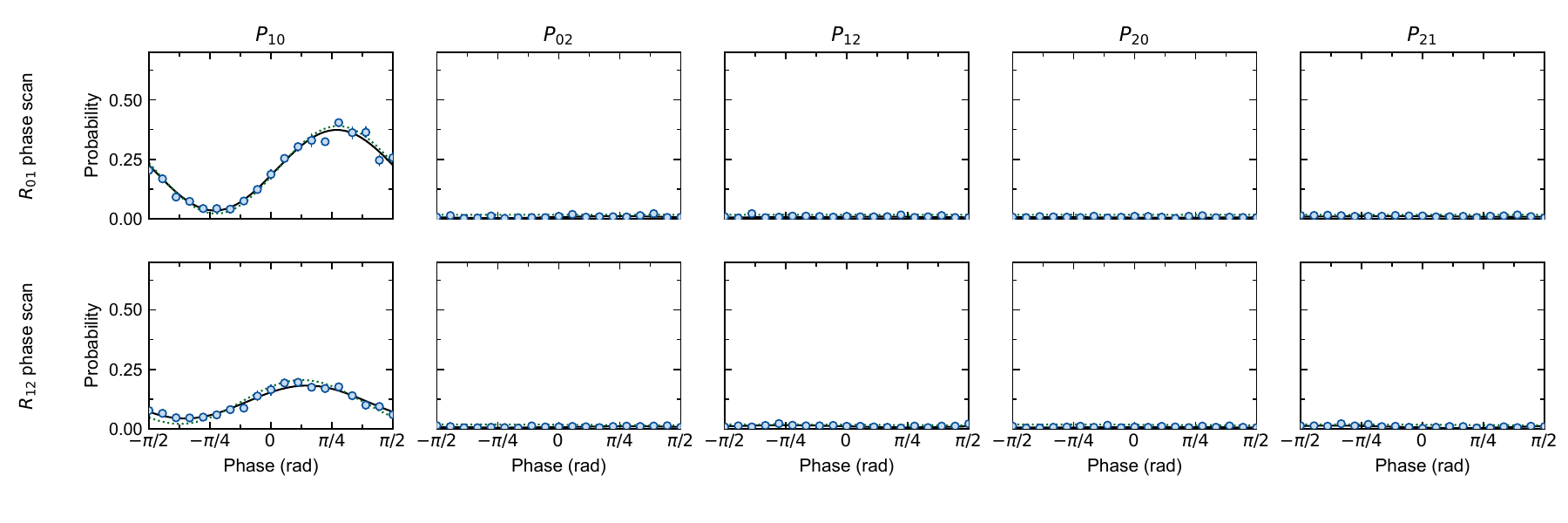}
\caption{Remaining outcome probabilities, apart from those shown in \cref{fig:3}(c), used to calibrate the nine-outcome likelihood model. The upper and lower rows show the $R_{01}$ and $R_{12}$ readout-phase scans. Points and error bars are the mean measured probabilities and their standard errors across atom pairs. Black curves are weighted least-squares fits of the cosine-sine model in \cref{eq:ScanResponseModel}, and green dashed curves are the circuit-model predictions from the independently characterized single-qutrit damping together with the two readout-phase offsets.}
\label{fig:OtherOutcomeProbabilities}
\end{figure*}

\cref{fig:3}(c) displays four of the nine computational-basis outcomes, the remaining five are shown in \cref{fig:OtherOutcomeProbabilities}.

\section{Maximum likelihood estimator and bootstrapping}
\label{sec:MLEAndBootstrap}

For the observed category counts $N_{mn}$, where $m$ and $n$ denote the two resolved qutrit outcomes, the number of outcomes retained after postselection on no atom loss is $N_{\mathrm{shot}}=\sum_{m,n}N_{mn}=29594$ (\SI{14.5}{\percent} discarded). The log likelihood for $\boldsymbol{\theta}=(\theta_1,\theta_2)^{\mathsf T}$ and its maximum-likelihood estimator are
\begin{equation}
\begin{aligned}
\ell(\boldsymbol{\theta}) &= \sum_{m,n=0}^{2}N_{mn}\log p_{mn}(\boldsymbol{\theta}),\\
\widehat{\boldsymbol{\theta}}_{\mathrm{MLE}} &= \underset{\boldsymbol{\theta}}{\arg \max} \  \ell(\boldsymbol{\theta}).
\end{aligned}
\label{eq:MLEDefinition}
\end{equation}

This estimator uses all correlations among the nine two-qutrit outcomes, rather than reducing the data to two independent Ramsey signals. The experimentally reported precision is obtained by applying this estimator to resampled or batched data and comparing the resulting covariance with the Fisher-information prediction.


The statistical uncertainty of $\widehat{\boldsymbol{\theta}}_{\mathrm{MLE}}$ is estimated by a nonparametric bootstrap of the data recorded at fixed phases. This dataset contains $N_{\mathrm{shot}}$ resolved pair outcomes, where $N_{\mathrm{shot}}$ is their total number and each outcome $(m,n)$ is the joint readout result of one atom pair. Consistent with the multinomial likelihood in \cref{eq:MLEDefinition}, each resolved pair outcome is treated as one observation and used as one resampling unit. Each bootstrap replicate draws $N_{\mathrm{shot}}$ outcomes with replacement, recomputes the counts $N_{mn}$, and repeats the maximization using the probability model in \cref{eq:ScanProbabilityModel}, yielding $\widehat{\boldsymbol{\theta}}_r^{*}$. A total of $20000$ replicas are generated and divided into twenty nonoverlapping blocks of $B=1000$ estimates for the statistics below.

The sample mean and covariance within a block are
\begin{equation}
\begin{aligned}
\overline{\boldsymbol{\theta}}^{*} &= \frac{1}{B}\sum_{r=1}^{B}
\widehat{\boldsymbol{\theta}}_r^{*},\\
\widehat\Gamma^{*} &= \frac{1}{B-1}\sum_{r=1}^{B}
\left(\widehat{\boldsymbol{\theta}}_r^{*}-\overline{\boldsymbol{\theta}}^{*}\right)
\left(\widehat{\boldsymbol{\theta}}_r^{*}-\overline{\boldsymbol{\theta}}^{*}\right)^{\mathsf T}.
\end{aligned}
\label{eq:BootstrapSampleStatistics}
\end{equation}
The plotted experimental point uses the first block: the horizontal coordinate for parameter $i$ is $\overline{\theta}_i^{*}$, the corresponding $1\sigma$ uncertainty is the sample standard deviation $\sqrt{(\widehat\Gamma^{*})_{ii}}$, and the vertical coordinate is the cost in \cref{eq:CostDefinition} for the estimate from all $N_{\mathrm{shot}}$ outcomes, $Y=\operatorname{Tr} (\widehat\Gamma^{*})$, multiplied by $\nu=N_{\mathrm{shot}}$ to clear the factor $1/\nu$ in \cref{eq:QuantumCRBCost}.

The vertical uncertainty is evaluated from all twenty blocks. If $\widehat\Gamma_b^{*}$ is the sample covariance in block $b$, the vertical $1\sigma$ error bar is the sample standard deviation $\operatorname{SD}_{b=1,\ldots,20}\left[N_{\mathrm{shot}}\operatorname{Tr} (\widehat\Gamma_b^{*})\right]$. This gives $N_{\mathrm{shot}}Y=0.82\pm 0.03$.



\section{$N$-qutrit extension}

To extend our analysis to $N$ qutrit systems, we consider a family of GHZ-like probe states
\begin{equation}
\ket{\psi_N}
=\sqrt{p_0}\ket{0}^{\otimes N}
+\sqrt{p_1}\ket{1}^{\otimes N}
+\sqrt{p_2}\ket{2}^{\otimes N},
\label{eq:si-ghz-family}
\end{equation}
where $p_0+p_1+p_2=1$.
Only the populations $p_j$ enter the quantum Fisher information matrix, because the generators are diagonal. 
On a computational-basis string $\boldsymbol{x}\in\{0,1,2\}^{N}$ the collective generators take the value $\sum_{n=1}^{N}\boldsymbol{g}_{x_n}$, where $\boldsymbol{g}_j=(\bra{j}g_1\ket{j},\bra{j}g_2\ket{j})^{\mathsf T}$, so their covariance matrix $\Gamma_{\rho}$ in an arbitrary $N$-qutrit probe $\rho$, defined as the mixed-state version of \cref{eq:si-generator-covariance}, depends only on the diagonal elements $\rho_{\boldsymbol{x}\boldsymbol{x}}=\bra{\boldsymbol{x}}\rho\ket{\boldsymbol{x}}$. We compare $\rho$ with the GHZ-like state whose weights are the mean level populations of $\rho$,
\begin{equation}
\begin{aligned}
p_j &= \frac{1}{N}\sum_{\boldsymbol{x}}\rho_{\boldsymbol{x}\boldsymbol{x}} n_j(\boldsymbol{x}),\\
\overline{\boldsymbol{g}} &= \sum_{j=0}^{2}p_j\boldsymbol{g}_j ,
\end{aligned}
\label{eq:si-ghz-populations}
\end{equation}
where $n_j(\boldsymbol{x})$ is the number of atoms in level $j$. Then one obtains
\begin{equation}
\begin{aligned}
{\cal F}_{\rho}
&\leq 4\Gamma_{\rho}\\
&= 4\sum_{\boldsymbol{x}}\rho_{\boldsymbol{x}\boldsymbol{x}}
\left[\sum_{n=1}^{N}\left(\boldsymbol{g}_{x_n}-\overline{\boldsymbol{g}}\right)\right]
\left[\sum_{m=1}^{N}\left(\boldsymbol{g}_{x_m}-\overline{\boldsymbol{g}}\right)\right]^{\mathsf T}\\
&= 4N\sum_{\boldsymbol{x}}\rho_{\boldsymbol{x}\boldsymbol{x}}\sum_{n=1}^{N}
\left(\boldsymbol{g}_{x_n}-\overline{\boldsymbol{g}}\right)
\left(\boldsymbol{g}_{x_n}-\overline{\boldsymbol{g}}\right)^{\mathsf T}\\
&\quad -4\sum_{\boldsymbol{x}}\rho_{\boldsymbol{x}\boldsymbol{x}}\sum_{n<m}
\left(\boldsymbol{g}_{x_n}-\boldsymbol{g}_{x_m}\right)
\left(\boldsymbol{g}_{x_n}-\boldsymbol{g}_{x_m}\right)^{\mathsf T}\\
&\leq 4N\sum_{\boldsymbol{x}}\rho_{\boldsymbol{x}\boldsymbol{x}}\sum_{n=1}^{N}
\left(\boldsymbol{g}_{x_n}-\overline{\boldsymbol{g}}\right)
\left(\boldsymbol{g}_{x_n}-\overline{\boldsymbol{g}}\right)^{\mathsf T}\\
&= 4N^2\sum_{j=0}^{2}p_j
\left(\boldsymbol{g}_{j}-\overline{\boldsymbol{g}}\right)
\left(\boldsymbol{g}_{j}-\overline{\boldsymbol{g}}\right)^{\mathsf T}\\
&= 4\Gamma_{\psi_N}\\
&= {\cal F}_{\psi_N} .
\end{aligned}
\label{eq:si-ghz-optimality}
\end{equation}
Therefore $\operatorname{Tr}[{\cal F}_{\rho}^{-1}]\geq \operatorname{Tr}[{\cal F}_{\psi_N}^{-1}]$ holds for any $\rho$, which means it is enough to optimize the weights $p_j$ for obtaining the optimal probe state.

For this family \cref{eq:si-ghz-family} one obtains
\begin{equation}
\operatorname{Tr}\left[{\cal F}^{-1}\right]
=\frac{1}{18N^2}
\left(
\frac{5}{p_0}
+\frac{2}{p_1}
+\frac{5}{p_2}
\right).
\label{eq:si-trace-qfim-family}
\end{equation}
Minimizing \cref{eq:si-trace-qfim-family} gives
\begin{equation}\label{eq:OptimalComponentWeights}
p_0=p_2=\frac{10-\sqrt{10}}{18},
\qquad
p_1=\frac{\sqrt{10}-1}{9} ,
\end{equation}
analogously to the case $N=2$.

The optimal probe is therefore
\begin{equation}\label{eq:OptimalNQutritProbe}
\ket{\Psi_N^{\mathrm{opt}}}=\sqrt{\frac{10-\sqrt{10}}{18}}
\left(\ket{0}^{\otimes N}+\ket{2}^{\otimes N}\right)
+\sqrt{\frac{\sqrt{10}-1}{9}}\ket{1}^{\otimes N}.
\end{equation}

The corresponding ideal joint-estimation cost is
\begin{equation}
Y_{\rm joint}^{\rm opt}
=
\operatorname{Tr}\left[{\cal F}^{-1}\right]_{\rm opt}
=
\frac{11+2\sqrt{10}}{9N^2}.
\label{eq:si-joint-cost}
\end{equation}

For equal allocation of atoms, the ideal individual-estimation benchmark is
\begin{equation}
Y_{\rm ind}
=
(\Delta\theta_1)^2+(\Delta\theta_2)^2
\geq
\frac{4}{N^2}.
\label{eq:si-individual-bound}
\end{equation}

The ideal joint- to individual-estimation ratio obtained from
\cref{eq:si-joint-cost,eq:si-individual-bound} is again
$Y_{\rm joint}/Y_{\rm ind}\approx 0.4812$, independent of $N$. 
\cref{fig:4} instead evaluates a scalable circuit family, which fixes an ordered pattern of global rotations and fully connected entangling blocks for every even $N$ at the cost of slightly suboptimal component weights in the probe state.

The fully connected entangling block is
\begin{equation}
C_N=\prod_{i<j}CZ_{ij},
\label{eq:FullyConnectedLayer}
\end{equation}
where $CZ_{ij}$ applies the controlled-$Z$ gate to atoms $i$ and $j$. 
In the sequences below, each $R_{01}$ and $R_{12}$ is the single-qutrit rotation of \cref{eq:R01MatrixForm,eq:R12MatrixForm} applied globally to all $N$ atoms. The preparation phases are
\begin{equation}
\begin{aligned}
\phi &\equiv -\frac{(4N-3)\pi}{4N} \pmod{\frac{2\pi}{N}},\\
\psi &\equiv \frac{(N-1)\pi}{2N} \pmod{\frac{2\pi}{N}}.
\end{aligned}
\label{eq:ScalablePreparationPhases}
\end{equation}
With the rightmost factor acting first, the preparation unitary is
\begin{equation}
\begin{aligned}
U_{\mathrm{prep}}^{(N)} ={}& R_{12}\left(-\frac{\pi}{2},\psi\right)\cdot C_N
\cdot R_{12}\left(\frac{\pi}{2},\psi+\frac{\pi}{2}\right)\\
&{}\cdot R_{01}\left(-\frac{\pi}{2},\phi\right)\cdot C_N
\cdot R_{01}\left(\frac{\pi}{2},\phi+\frac{\pi}{2}\right).
\end{aligned}
\label{eq:ScalablePreparationSequence}
\end{equation}
Applied to $\ket{0}^{\otimes N}$, this sequence prepares
\begin{equation}
\ket{\phi_N} = \frac{1}{\sqrt{2}}\ket{0}^{\otimes N}
+\frac{1}{2}\ket{1}^{\otimes N}
+\frac{1}{2}\ket{2}^{\otimes N}.
\label{eq:ScalableProbe}
\end{equation}
Under the encoding in \cref{eq:CollectiveGeneratorsAndEncoding}, the encoded probe state is
\begin{equation}
\begin{aligned}
\ket{\phi_N(\boldsymbol{\theta})} &= \frac{1}{\sqrt{2}}e^{-iN\theta_1/\sqrt{2}}\ket{0}^{\otimes N}
+\frac{1}{2}e^{iN(\theta_1-\theta_2)/\sqrt{2}}\ket{1}^{\otimes N}\\
&\quad +\frac{1}{2}e^{iN\theta_2/\sqrt{2}}\ket{2}^{\otimes N}.
\end{aligned}
\label{eq:ScalableEncodedProbe}
\end{equation}

The joint measurement applies the readout unitary
\begin{equation}
V_{\mathrm{ro}}^{(N)} = R_{12}\left(\frac{\pi}{2},\frac{\pi}{2N}\right)
\cdot R_{01}(\alpha_{\rm mid},\phi_{\rm mid})
\cdot C_N\cdot R_{12}\left(\frac{\pi}{2},\frac{\pi}{2N}\right),
\label{eq:ScalableJointReadout}
\end{equation}
where the angle and phase of the central $R_{01}$ pulse are
\begin{equation}
\begin{aligned}
&(\alpha_{\rm mid},\phi_{\rm mid}) =\\
&\begin{cases}
\left(2\arccos \dfrac{1}{\sqrt{3}},\dfrac{7\pi}{4N}\right),
& N \equiv 0\pmod 4,\\[6pt]
\left(2\pi-2\arccos \dfrac{1}{\sqrt{3}},\dfrac{\pi}{4N}\right),
& N \equiv 2\pmod 4.
\end{cases}
\end{aligned}
\label{eq:ScalableReadoutParameters}
\end{equation}
The readout is followed by a measurement in the computational basis. The density matrix immediately before this measurement and the probability of a full computational-basis outcome $\boldsymbol{x}\in \{0,1,2\}^N$ are
\begin{equation}
\begin{aligned}
\rho_{\mathrm{out}}^{\rm joint}(\boldsymbol{\theta}) &= V_{\mathrm{ro}}^{(N)}\ket{\phi_N(\boldsymbol{\theta})}\bra{\phi_N(\boldsymbol{\theta})}V_{\mathrm{ro}}^{(N)\dagger},\\
p_{\boldsymbol{x}}(\boldsymbol{\theta}) &= \langle\boldsymbol{x}|\rho_{\mathrm{out}}^{\rm joint}(\boldsymbol{\theta})|\boldsymbol{x}\rangle.
\end{aligned}
\label{eq:ScalableFinalStateProbabilities}
\end{equation}

Thus the measurement probabilities are the diagonal elements of the final density matrix. Since the final state is invariant under particle permutations, computational-basis strings with identical occupation counts $(n_0,n_1,n_2)$ have equal probabilities. If $p_{n_0n_1n_2}(\boldsymbol{\theta})$ is the probability of any one string with those counts, the classical Fisher information matrix is
\begin{equation}
\begin{aligned}
&F_{\mu\nu}(\boldsymbol{\theta}) \\
&= \sum_{n_0+n_1+n_2=N} \frac{N!}{n_0!n_1!n_2!}
\frac{\partial_{\theta_\mu}p_{n_0n_1n_2}(\boldsymbol{\theta})
\partial_{\theta_\nu}p_{n_0n_1n_2}(\boldsymbol{\theta})}
{p_{n_0n_1n_2}(\boldsymbol{\theta})}.
\end{aligned}
\label{eq:CountClassCFIM}
\end{equation}

At $\boldsymbol{\theta}=0$, the diagonal probabilities depend only on the residue class $r(\boldsymbol{x})\equiv n_0(\boldsymbol{x})+2n_2(\boldsymbol{x})+N/2\pmod 3$. Each of the three residue classes contains $3^{N-1}$ strings. Evaluating the diagonal probabilities in \cref{eq:ScalableFinalStateProbabilities} and substituting them into \cref{eq:CountClassCFIM} gives
\begin{equation}
\begin{aligned}
F_{\rm joint}(0) &= N^2
\begin{pmatrix}
11/8 & -1/2\\
-1/2 & 1
\end{pmatrix},\\
 \operatorname{Tr} \left[F_{\rm joint}(0)^{-1}\right] &= \frac{19}{9N^2}.
\end{aligned}
\label{eq:ScalableReadoutSaturation}
\end{equation}
Applying \cref{eq:PureStateQFIM} to \cref{eq:ScalableProbe} gives the same matrix, so the readout in \cref{eq:ScalableJointReadout} saturates the quantum Cram\'er-Rao bound of this probe locally at $\boldsymbol{\theta}=0$.

Individual estimation assigns $N/2$ atoms to each parameter. For $ab\in \{01,12\}$, the preparation and readout unitaries are
\begin{equation}
\begin{aligned}
U_{\mathrm{ind}}^{(ab)} &= R_{ab}\left(\frac{\pi}{2},\frac{\pi}{N}+\frac{\pi}{2}\right)
\cdot C_{N/2}\cdot R_{ab}\left(\frac{\pi}{2},\frac{\pi}{N}\right),\\
V_{\mathrm{ind}}^{(ab)} &= R_{ab}\left(\frac{\pi}{2},\frac{3\pi}{2}+\frac{\pi}{N}\right).
\end{aligned}
\label{eq:IndividualProtocol}
\end{equation}
The subensemble assigned to the $01$ protocol begins in $\ket{0}^{\otimes N/2}$, and that assigned to the $12$ protocol begins in $\ket{1}^{\otimes N/2}$. At $\boldsymbol{\theta}=0$, the computational-basis probabilities generated by either readout give
\begin{equation}
\begin{aligned}
F_{\rm ind}(0) &= \frac{N^2}{2}I_2,\\
 \operatorname{Tr} \left[F_{\rm ind}(0)^{-1}\right] &= \frac{4}{N^2}.
\end{aligned}
\label{eq:IndividualReadoutBound}
\end{equation}
The ratio plotted in \cref{fig:4} is
\begin{equation}
\frac{Y_{\rm joint}}{Y_{\rm ind}} = \frac{ \operatorname{Tr} (F_{\rm joint}^{-1})}{\operatorname{Tr} (F_{\rm ind}^{-1})}.
\label{eq:FigureFourRatio}
\end{equation}
For noiseless circuits, \cref{eq:ScalableReadoutSaturation,eq:IndividualReadoutBound} give $Y_{\rm joint}/Y_{\rm ind}=19/36\approx 0.5278$. This value is slightly higher than $0.4812$ because the scalable probe uses component weights $(1/2,1/4,1/4)$ rather than the optimal weights in \cref{eq:OptimalComponentWeights}.

For the noisy curves in \cref{fig:4}, each $R_{ab}$ pulse is followed by an effective map that contracts the Bloch vector on the two-level subspace spanned by $\ket{a}$ and $\ket{b}$ by a factor $\eta_R$ and retains the remaining qutrit matrix elements. On the two-level block $\rho_{ab}$, the map is
\begin{equation}
\mathcal E_{R,\eta_R}^{(ab)}(\rho_{ab}) = \eta_R\rho_{ab}
+(1-\eta_R)\frac{\operatorname{Tr} (\rho_{ab})}{2}I_{ab},
\label{eq:RPulseEffectiveNoise}
\end{equation}
where $I_{ab}$ is the identity on this subspace. Each pairwise CZ gate is followed by local dephasing on its two atoms. For either atom,
\begin{equation}\label{eq:SelectiveDephasing}
\mathcal D_{1,\eta_C}\left(\ket{\mu}\bra{\lambda}\right) = \eta_C^{(\delta_{\mu,1}-\delta_{\lambda,1})^2}\ket{\mu}\bra{\lambda},
\end{equation}
where $\mu,\lambda\in \{0,1,2\}$ and $\eta_C$ is the coherence factor associated with one pairwise gate. The rotation and pairwise-gate infidelities used for the curves are
\begin{equation}
\begin{aligned}
\epsilon_R &= \frac{1-\eta_R}{2},\\
\epsilon_{CZ} &= \frac{4}{45}\left(7-5\eta_C-2\eta_C^2\right).
\end{aligned}
\label{eq:GateInfidelityParameters}
\end{equation}
Here $\epsilon_R$ is the average infidelity within the two-level subspace, and $\epsilon_{CZ}$ is the average infidelity of one pairwise CZ gate.

Replacing each ideal gate in the preparation and readout sequences by these noisy operations produces the final joint and individual density matrices. Their computational-basis diagonal elements are grouped by occupation counts and inserted into \cref{eq:CountClassCFIM}. The two single-parameter Fisher informations form the diagonal of $F_{\rm ind}$, and \cref{eq:FigureFourRatio} is then evaluated at $\boldsymbol{\theta}=0$. The plotted curves set $\epsilon_R=\epsilon_{CZ}$ to $10^{-3}$, $10^{-4}$, $10^{-5}$, and $0$. Since every $C_N$ layer contains $\binom{N}{2}$ pairwise gates, the accumulated dephasing increases with $N$ and eventually eliminates the joint-estimation advantage for nonzero infidelity.

\subsection{White-noise robustness}

The model underlying \cref{fig:4} assigns an error to each gate of the circuit. 
We now apply the same white-noise model \cref{eq:ReadoutDesignWhiteNoiseModel} to the optimal $N$-qutrit probe. This determines what fraction of the target probe state must be retained for a joint-estimation advantage at each $N$.
The depolarized $N$-qutrit probe is
\begin{equation}
\rho_{\eta}^{\rm joint}
=\eta\ket{\Psi_N^{\rm opt}}\bra{\Psi_N^{\rm opt}}
+(1-\eta)\frac{I_{3^N}}{3^N},
\label{eq:si-white-noise-model}
\end{equation}
where $\eta$ is the coherent-state visibility. For unitary phase encoding, the quantum Fisher information matrix of this depolarized pure-state family is reduced by the factor
\begin{equation}
\xi_N(\eta)
=\frac{\eta^2}{\eta+2(1-\eta)/3^N}.
\label{eq:si-white-noise-factor}
\end{equation}
Therefore
\begin{equation}
Y_{\rm joint}^{\eta}
=\frac{1}{\xi_N(\eta)}
\frac{11+2\sqrt{10}}{9N^2}.
\label{eq:si-noisy-joint-cost}
\end{equation}
Comparing \cref{eq:si-noisy-joint-cost} with the noiseless individual-estimation benchmark in \cref{eq:si-individual-bound}, noisy joint estimation remains advantageous whenever
\begin{equation}\label{eq:GlobalWhiteNoiseAdvantage}
\xi_N(\eta)
>\frac{11+2\sqrt{10}}{36}
\simeq 0.4812 .
\end{equation}
Because the depolarizing background is spread over a Hilbert space of dimension $3^N$, the robustness improves with atom number. In the large-$N$ limit, $\xi_N(\eta)\simeq \eta$, so the joint strategy remains advantageous for visibilities above approximately $0.4812$. 
The two models constrain different resources: \cref{eq:GlobalWhiteNoiseAdvantage} sets the visibility required of the probe state, which relaxes towards $0.4812$ as $N$ grows, while \cref{fig:4} sets the per-gate error budget of the circuit that prepares and reads it out.

\end{document}